\documentstyle[12pt,aasms4]{article}

\def\msun{$M_{\odot}$}
\def\Z{\rm Z}
\def\hii{H\thinspace{$\scriptstyle{\rm II}$}~}
\def\gsim{\mathrel{\rlap{\lower4pt\hbox{\hskip1pt$\sim$}}
    \raise1pt\hbox{$>$}}}         

\begin{document}

\title{Is High Primordial Deuterium Consistent \\
With Galactic Evolution?}

\author{Monica Tosi}
\affil{Osservatorio Astronomico di Bologna \\
Via Zamboni 33, 40126 Bologna, Italy}
\author{Gary Steigman}
\affil{Departments of Physics and Astronomy, The Ohio State University, \\
174 West 18th Avenue, Columbus, OH 43210, USA}
\author{Francesca Matteucci}
\affil{Dipartimento di Astronomia, Universit\`a di Trieste, SISSA, \\
Via Beirut 2-4, I-34013 Trieste, Italy}
\and
\author{Cristina Chiappini}
\affil{Dipartimento di Astronomia, Universit\`a di Trieste, SISSA, \\
Via Beirut 2-4, I-34013 Trieste, Italy \\
\& \\
Instituto Astron$\hat{\rm o}$mico e Geof\' isico, Universidade de 
S$\tilde{\rm a}$o Paulo \\
Av. Miguel Stefano 4200, S$\tilde{\rm a}$o Paulo, S.P. 04301-904, Brasil}



\begin{abstract}
Galactic destruction of primordial deuterium is inevitably linked through 
star formation to the chemical evolution of the Galaxy.  The relatively
high present gas content and low metallicity suggest only modest D-destruction.
In concert with deuterium abundances derived from solar system and/or
interstellar observations this suggests a primordial deuterium abundance
in possible conflict with data from some high-redshift, low-metallicity
QSO absorbers.  We have explored a variety of chemical evolution models
including infall of processed material and early, supernovae-driven winds
with the aim of identifying models with large D-destruction which are
consistent with the observations of stellar-produced heavy elements.  
When such models are confronted with data we reconfirm that only modest 
destruction of deuterium (less than a factor of 3) is permitted.  When 
combined with solar system and interstellar data these results favor the 
low deuterium abundances derived for the QSO absorbers observed by Tytler 
et al. (1996) and by Burles \& Tytler (1996).
\end{abstract}


\section{Introduction}

Observations of high-redshift absorbers along the lines-of-sight 
to distant QSOs reveal nearly unprocessed primordial gas.  In
a handful of such systems there is evidence for deuterium; the
inferred abundance may be very nearly the primordial value.  For
two high-z, low-Z systems Tytler and collaborators (Tytler, Fan \&
Burles 1996, Burles \& Tytler 1996) derive low deuterium abundances, 
D/H = (2.4$\pm0.5)\times10^{-5}$, similar to those inferred from
observations of the local interstellar medium (Linsky et al. 1993, Linsky 
1998).  Such low values have been challenged by Songaila et al. (1997), 
who suggest a larger deuterium abundance, D/H $>4\times10^{-5}$, for 
one of the Burles \& Tytler (1996) absorbers.  In contrast, deuterium 
abundances nearly an order of magnitude higher, D/H = (19$\pm4)\times10^{-5}$, 
have been claimed by Carswell et al. (1994), Songaila et al. (1994) and 
by Rugers \& Hogan (1996) for other high-redshift systems with metal 
abundances equally close to primordial.  However, using better data 
obtained for one of these systems, Tytler, Burles \& Kirkman (1997) 
find no evidence for deuterium and exclude the published (high) abundance 
at the ``10$\sigma$" level.  Nonetheless, Hogan (1997 and Private 
Communication), Wampler (1996 and Private Communication) and Songaila 
(1997) argue that the spectra of other absorbing systems require high 
D/H (e.g., Webb et al. 1997).  The high-D, low-D dispute awaits more 
data for its resolution.

Even though these QSO absorbers have very low metal abundances, 
suggesting very little stellar activity, their D content must 
still be considered only a lower limit to the primordial deuterium 
abundance, (D/H)$_P$, since D is only destroyed by stellar processes 
subsequent to Big Bang Nucleosynthesis (BBN).  To reconcile the current 
low abundance of D inferred from observations of the local interstellar 
medium (ISM), (D/H)$_{ISM} = (1.6\pm0.1)\times10^{-5}$ (Linsky et al. 
1993), with a primordial value as large as that suggested by Hogan and 
collaborators requires that deuterium has been destroyed by an order 
of magnitude, or more, in the course of the evolution of the Galaxy.  
Although such high abundances are now strongly challenged observationally, 
it is still important to investigate whether any chemical evolution 
scenarios exist which permit large D-destruction factors and, if so, 
how their predictions compare with the available observational constraints 
on the distribution -- in space and in time -- of the abundances of the 
other elements.

\section{Galactic Evolution of Deuterium}

As summarized by Tosi (1996 and references therein), those models 
for the evolution of the disk of our Galaxy which are in agreement 
with the largest set of available observational constraints predict 
that deuterium has been depleted by less than a factor of three.  
Thus, to reproduce the abundances, by mass, observed in the solar 
system, X$_{2\odot}=(3.6\pm 1.3)\times 10^{-5}$ and in the local 
interstellar medium, X$_{2ISM}=(2.2\pm 0.3)\times 10^{-5}$ (Geiss 
1993; Linsky et al. 1993, respectively), none of these {\it good, 
standard} models would allow initial deuterium abundances higher 
than X$_{2i}\simeq 8\times 10^{-5}$.  Such a low upper bound to 
primordial D is much lower than the value, X$_{2RH}=(29\pm 6)\times 
10^{-5}$, suggested by Rugers \& Hogan (1996, hereinafter RH).  In 
all the models considered by Tosi (1996): Carigi (1996), Chiappini 
\& Matteucci (1996), Dearborn, Steigman \& Tosi (1996, hereinafter DST), 
Galli et al. (1995), Prantzos (1996), those with large initial deuterium 
either cannot fit the D-abundances observed in the solar system and/or 
in the local ISM, or cannot account for the presently observed amount 
of disk gas and/or the abundances of the heavier elements.  Although 
other models designed to permit higher D-destruction have been proposed 
(e.g. Vangioni-Flam \& Audouze 1988; Vangioni-Flam, Olive \& Prantzos 
1994; Olive et al. 1995; Scully et al. 1996, 1997), in general they have 
not always been tested against all the available observational constraints 
and their self-consistency has been questioned (e.g. Edmunds 1994, Prantzos 
1996).

To investigate if it is possible to avoid this apparent inconsistency 
(high primordial D, low Galactic D), we have pursued two distinct 
approaches. On the one hand, we have developed a series of general 
(i.e., model-independent) arguments along the line of reasoning described 
by Steigman \& Tosi (1995), and, on the other hand, we have actually 
computed the evolution predicted for several specific, albeit {\it ad 
hoc}, models for the halo and the disk of the Galaxy which were designed 
to maximize D-destruction.

\section{A Model-Independent Approach}

The evolution of deuterium after BBN is straightforward since, when 
incorporated in a star, it is completely destroyed, burned to $^3$He 
during the pre-main sequence evolution.  If the ``virgin" fraction of 
the ISM (either now or at the time of the formation of the solar system) 
were known, the primordial (pre-Galactic) D-abundance could be inferred 
directly from ISM or solar system observations.  The more material cycled 
through stars, the more D-destruction, and vice-versa.  But, the more
material cycled through stars, the more heavy elements are synthesized 
and the more mass is tied up in unevolved low mass stars and in stellar 
remnants.  Efficient D-destruction and efficient star formation go 
hand-in-hand.  Thus, there is the possibility that the observed metallicity 
and/or the disk gas-fraction may be exploited to constrain the deuterium 
depletion factor in a manner insensitive to the details of specific models
of Galactic evolution.  We explore one such approach below; see Steigman 
\& Tosi (1995) for an alternate approach.

Consider a ``representative" sample of the ISM.  Some fraction (by mass) 
of that material, $f_0$, will have never been through stars.  In this 
``virgin" fraction the deuterium is primordial (X$_2 =$ X$_{2P}$), and 
there are no metals (Z = 0).  The remainder of the material will have 
been through one or more generations of stars which destroyed all the 
deuterium (X$_2$ = 0), but which did produce some heavy elements.  For 
example, in a fraction $f_1$, which has been through one (and only one) 
generation of stars, there will be no deuterium but there are some heavy 
elements.  Estimating the metallicity of this ``one-generation" fraction 
is complicated.  Some of the matter may be from long-lived, low-mass stars 
formed long ago and some may be from short-lived, high-mass stars formed
very recently.  Thus, the gas returned to the ISM will have come from a 
mixture of stars of all masses, including the very low-mass ones which 
destroy deuterium but do not contribute much to the enhancement of the 
heavy element abundances and the higher-mass stars which destroy deuterium
and do enhance the Galactic metallicity.  In this manner the average 
metallicity ($<\Z_1>$) of this one-generation sub-sample is a mean, 
weighted by the initial mass function (IMF).

Generalizing the above description to material which has been through 
2, 3, ... generations of stars ($f_2$, $f_3$, ...), we may relate the 
deuterium and heavy element mass fractions as follows:

\begin{equation}
X_2 = f_0X_{2P} 
\end{equation}
\begin{equation}
\Z = f_1<\Z_1> + f_2<\Z_2> + f_3<\Z_3> + .... 
\end{equation}
\begin{equation}
1 = f_0 + f_1 + f_2 + f_3 + .... 
\end{equation}
As they stand, these relations are not very useful.  However, if there exists
a lower bound ($\Z_{min}$) to all the $<\Z_i>$ ($\Z_{min}$ $\leq$ $<\Z_i>$), 
eq. (2) can be rewritten as:

\begin{equation}
\Z \geq (1 - f_0) \Z_{min} 
\end{equation}
or,
\begin{equation}
f_0 \geq 1 - \Z/\Z_{min} 
\end{equation}
so that,
\begin{equation}
X_{2P} \leq X_2/(1 - \Z/\Z_{min}). 
\end{equation}

Equations (4-6) summarize, in a compact form, the anticipated connections: 
the more material cycled through stars, the higher will be the metallicity 
and the more will deuterium have been destroyed.  Thus, the present (ISM 
or presolar) value of $\Z$, supplemented with an estimate of $\Z_{min}$, 
may be used to provide a bound on $f_0$ and, therefore, on X$_{2P}$.  To 
implement this approach, we have chosen oxygen as a good tracer of the 
total metallicity, and to estimate $\Z_{min}$ we have computed the 
contribution to the metallicity from a single stellar population; 
$\Z_{min}$ $\equiv$ min(${<\Z_1>}$), where:

\begin{equation}
<\Z_1> = 
\int_{m_l}^{m_u}M_{16,net}\phi(m)dm  / \int_{m_l}^{m_u}M_{ej}\phi(m)dm. 
\end{equation}
$M_{ej}$ is the total mass ejected by stars of mass $m$, $M_{16,net}$
is the mass ejected in the form of newly synthesized oxygen and $\phi(m)$ 
is the IMF.  The oxygen enrichment provided by one generation of stars 
thus depends on the stellar nucleosynthesis yields we have adopted, 
and on the IMF through its slope ($\alpha$) and upper ($m_u$) and lower 
($m_l$) mass limits. 

To explore the uncertainties traceable to stellar yields we have computed 
$<\Z_1>$ adopting both the minimum and the maximum oxygen yields from the 
literature (Chiosi \& Caimmi 1979, hereinafter CC79, and Woosley \& Weaver 
1995, WW95, respectively).  Similarly, for the IMF we have chosen the two 
most extreme popular slopes appropriate for the range of massive stars 
which produce oxygen: $\alpha$ = 3.3 as in Tinsley (1980) and $\alpha$ = 
2.35 as in the extrapolation of the Salpeter IMF (1955); we allow both 
$m_u$ and $m_l$ to vary.  The top panel of Figure~\ref{imf} shows 
the behaviour of the ``virgin" fraction, $f_0 = 1-\Z_{16}/<\Z_{16}>$, 
as a function of the lower mass limit ($m_l$) to the IMF (with $m_u$ 
= 100\msun), for oxygen yields taken from WW95 (dotted line) and from 
CC79 (solid line) respectively.  The IMF is Tinsley's (1980) and we 
adopt for the local oxygen abundance at the time the solar system 
formed: $\Z_{16\odot} = 9.5\times 10^{-3}$ (Anders \& Grevesse 1989).  
We have explored all combinations of the parameters and find for the 
minimum of the one-generation yield, $<\Z_1>_{min}$ = 0.012.  Using 
this value for $\Z_{min}$ we conclude that the fraction of gas which 
has never been through stars is $f_0>0.21$.  This suggests (see eq. 1) 
that under ``normal" conditions (standard yields, reasonable IMFs), 
deuterium cannot be destroyed by more than a factor of 5.

This upper bound to D-destruction is clearly sensitive to our choices of 
$<\Z_1>_{min}$ and of $\Z_{\odot}$. To obtain an extreme upper bound we 
have been careful to adopt those values for both quantities which permit 
the maximum D-depletion.  The Anders \& Grevesse (1989) estimate of the 
solar metallicity is quite robust.  Although the Sun may not provide a fair 
sample of the local ISM 4.5 Gyr ago, its abundance may, instead, provide an 
upper bound to the metallicity of the local ISM at that epoch since the Sun 
appears to be more metal-rich than the average of the local stars of its 
age.  Indeed, if the average local metallicity when the solar nebula collapsed 
were similar to, or even lower than that derived from Orion and the local 
\hii regions, i.e., 0.2 dex lower than $\Z_{\odot}$, our upper bound on 
D-destruction will be reduced.  As for $<\Z_1>_{min}$, our bound is derived 
from those theoretical yields in the literature with the lowest O-production  
and using the IMF for the local region with smallest fraction of massive stars 
(i.e., of oxygen producers).  For these reasons we regard a factor of five 
as a robust upper bound to possible D-destruction.

Our bound on D-destruction is general in the sense that it is independent 
of the details of specific chemical evolution models; the only inputs are 
from the IMF and the stellar yields.  However, it is key that the 
amount of deuterium destruction and the metallicity are linked through 
our assumption that all the stellar ejecta do contribute to the enrichment 
of the ISM.  This will be the case in the absence of winds capable of 
removing from the Galaxy part or all of the stellar ejecta.  In contrast, 
this D-destruction/metallicity connection may be avoided if winds, perhaps 
triggered by the explosions of multiple supernovae as originally suggested 
by Larson (1974) for elliptical galaxies, were present earlier in the 
evolution of the Galaxy.  In the extreme case of the total removal 
(permanently!) from the Galaxy of all the ejecta of stars more massive 
than a given mass, the effect of such a wind on the oxygen enrichment 
in the ISM is equivalent to cutting off the IMF at a value of $m_u$ equal 
to that mass.  Such an effect is illustrated in the bottom panel of 
Figure~\ref{imf} for two choices of $m_l$ (0.8\msun and 5\msun).  For the 
WW95 yields there remains a very tight constraint from the connection 
between metallicity and D-destruction, while the CC79 yields permit much 
more destruction.

As noted by Pagel (Private Communication), there is another way in 
which the D-destruction/metallicity connection may be weakened: black 
hole formation.  If all stars with $m \geq m_{\rm bh}$ collapse without
returning any newly synthesized material to the ISM, the average yield
per stellar generation will be reduced.  The effect of this, too, may 
be inferred from the bottom panel of Figure~\ref{imf}; from the point 
of view of chemical enrichment, it is equivalent to the material from 
such stars being expelled from the Galaxy via winds.  There are two 
potential problems associated with too small a value of $m_{\rm bh}$ 
or, equivalently, with winds which effectively cut off the IMF above 
$m_u$.  In both cases the oxygen yield is more strongly reduced than 
the helium yield so that the effect is to increase their ratio, 
$\Delta$Y/$\Delta$Z (Maeder 1992).  H\o g et al. (1998) and Thuan \& 
Izotov (1998) argue that $\Delta$Y/$\Delta$Z $\leq$ 3.  From Figure 15 
in Maeder (1992), for the Scalo (1986) IMF, we infer that to satisfy 
this limit requires $m_{\rm bh} \geq$ 55\msun.  Since for our ``robust" 
bound we use Tinsley's (1980) IMF which has fewer massive stars than 
Scalo's (1986), according to Maeder's (1992) arguments the lower limit 
to $m_{\rm bh}$ must be even larger.  As a result, (assuming $m_u = 
m_{\rm bh}$) even with the CC79 yields there is a limit on $f_0$ (see 
Figure~\ref{imf}).  Thus, depending on the actual yields (CC79 or WW95) 
and on the appropriate lower mass limit $m_l$, limits to $\Delta$Y/$\Delta$Z 
provide bounds on deuterium destruction in the presence of winds and/or 
black hole formation.  Another constraint on mass loss or black hole 
formation comes from the observed gas fraction which will decrease if 
significant mass is removed from the ISM, leaving more mass locked into 
long-lived stars and/or stellar remnants (Edmunds 1994).  Thus, although 
winds and/or black hole formation are promising avenues to pursue if the 
D-destruction/metallicity connection is to be weakened, there are still 
some observational constraints which must be satisfied.

These general results serve as a guide, permitting us to identify 
potentially fruitful approaches to designing chemical evolution models 
which, while destroying significant amounts of deuterium, may avoid 
exhausting the interstellar gas and/or overproducing the heavy elements.  
With standard IMFs and no significant winds, large D-destruction requires 
that so much mass has been cycled through stars that the observed metallicity 
would be exceeded.  This D-destruction/metallicity connection may be broken, 
for example, by adopting a nonstandard IMF biased against high-mass stars 
which are the main oxygen producers.  Although this will allow more gas to 
be cycled through stars, destroying deuterium while avoiding overproducing 
oxygen, at present there is currently no significant evidence for an anomalous 
IMF (Wyse 1997).  Alternatively, as suggested by Vangioni-Flam \& Cass\'e 
(1995) and by Scully et al. (1997), the offending metals might have been 
removed by strong winds.  Since there is no good evidence for variations 
of the IMF (e.g. Richer \& Fahlman 1996, Wyse 1997) we will pursue the 
latter approach in our search for specific Galactic evolution models 
which may be capable of destroying deuterium efficiently while preserving 
consistency with the observational data.
 
\section{Chemical Evolution Models}

To explore in detail whether Galactic winds and/or infall of D-depleted
gas may lead to observationally consistent models which permit high 
D-destruction, we have computed a series of numerical chemical evolution 
models including such winds.  We describe these models below and compare 
their predictions with the observational data available for our Galaxy.  
In following the evolution of the elements produced mainly by long-lived 
stars, such as $^3$He, $^{14}$N and $^{56}$Fe, it is of fundamental importance 
that the instantaneous recycling approximation is avoided.  

Bearing in mind that the initial disk abundance, X$_{2i}$, may not 
necessarily be the primordial one, X$_{2P}$, we consider two possible 
scenarios:

a) For the {\it good, standard} models summarized by Tosi (1996), fitting 
the observed ISM D-abundance constrains the initial disk deuterium abundance
to be X$_{2i}$ $\leq6\times 10^{-5}$, in contrast to the RH primordial
value of X$_{2P}$ = X$_{2RH}=29\times 10^{-5}$.  In this case most of the 
D-destruction must have occurred prior to disk formation.  Therefore, in 
this case the disk must form out of already processed gas.  This requires 
either that an unknown (and improbable) mechanism burned most of the 
primordial D prior to galaxy formation (without producing a significant 
metallicity), or that most of the destruction must have taken place during 
the halo phase.  This latter hypothesis assumes that the disk formed out
of gas shed from the halo which is in disagreement with recent observational
results (Ibata \& Gilmore 1995; see the discussions in Chiappini et al. 1997a
and in Pagel \& Tautvaisiene 1995).

b) If the initial disk abundance is, instead, X$_{2i}=$X$_{2P}=29\times 
10^{-5}$, it is necessary to invoke a much larger D consumption during 
disk evolution to deplete the deuterium abundance down to the observed 
solar system and local ISM values.  This, however, requires a much larger 
early star formation rate (SFR) which, in turn, as we've seen in our 
general considerations above, would lead to excessive metal enrichment 
(see e.g. Vangioni-Flam \& Audouze 1988).  These conclusions may possibly 
be evaded if the evolution includes the effect of strong galactic winds 
(see e.g. Scully et al. 1997) and/or the infall of D-depleted gas.  In 
any case, a very high initial SFR in the disk violates the constraint 
that the peak SFR can only be a factor of a few larger than the present 
SFR (e.g., Twarog 1980, Prantzos 1998).

During the early halo phase, if the SFR were much higher, so too would
have been the rate of supernova explosions, perhaps triggering strong 
winds.  In contrast, strong winds are highly improbable at present in 
the disk of our Galaxy (except perhaps in its outer layers) where both 
the gravitational potential and the gas density are likely too high to 
allow the gas to reach escape velocity (Spitzer 1990, Tenorio-Tagle 1996).  
Fountains (i.e., gas leaving the disk temporarily but returning later 
somewhere in the disk) may occur, but they won't remove the metals 
permanently.  Before considering models with strong, early winds we 
first explore the effects of infall.

\subsection{Models With Infall}

Infall generally has the effect of raising the disk D-abundance towards
the primordial value.  It is difficult to avoid this conclusion since
it has been shown, both theoretically (Tosi 1988b, Matteucci \& Fran\c cois 
1989, hereinafter MF) and observationally (e.g. Savage \& de Boer 
1981) that the infalling gas has a metallicity of, at most, 0.2--0.3$\Z_
{\odot}$.  Therefore, it is quite unlikely that all the deuterium in the
infalling gas has been destroyed.  For the sake of completeness, however, 
we have revisited the broad range of models described in DST and in Tosi
(1988a), allowing for arbitrarily low D/H in the infalling gas.  The 
predictions for the local evolution of deuterium in one representative
set of these models (see DST and Tosi 1988a for the model assumptions 
and parameters), which displays the key trends common to all such models, 
are shown in Figure~\ref{dinf}, together with the 2$\sigma$ ranges of 
the abundances (by mass) derived from solar system and local ISM 
observations.  Also shown in the figure are two models from DST: the 
solid line corresponds to model 1-T-Vb [SFR = exp(-t/15Gyr); infall 
F = 4$\times 10^{-3}$\msun$kpc^{-2}yr^{-1}$; metallicity \Z$_{infall}=
0.2\Z_{\odot}$] and the dotted line to the no infall model (NI-Z-III).  
None of these models is capable of satisfactorily reconciling a large 
primordial abundance, X$_{2P}$ = X$_{2RH}$, with the observed pre-solar 
and local ISM values.  Indeed, to reach the present ISM value starting 
from such a high initial abundance requires nearly D-free infall, 
X$_{2,infall} = 0.1$X$_{2P}$ (long-dashed line), but then the D-abundance 
predicted for the solar system is larger than that observed.  For the 
extreme choice, X$_{2,infall}$ = 0 (short-dashed line), the solar 
abundance can be fitted, but then the predictions for the D-abundance 
at the present epoch are too low compared to those observed in the local 
ISM.  The best compromise is achieved by assuming X$_{2,infall}$ = 
0.2X$_{2P}$ (dash-dotted line) which, however, corresponds to a maximum 
allowed primordial abundance, X$_{2P}=20\times 10^{-5}$ ((D/H)$_P = 
13\times 10^{-5}$), which, while large, is still lower than the RH 
value by a factor of 1.5.  Even this case is not entirely realistic 
since {\it standard} models (cf. Tosi 1996) suggest that the metallicity
will exceed solar in gas in which deuterium has been depleted by a
factor of 5.  Therefore D-depleted infalling gas will have high 
metallicity, while deuterium should be virtually undepleted in low
metallicity infalling gas.

The no-infall (NI) model (dotted line) actually predicts {\it less} 
D-destruction than even those models with infall of D-depleted gas.  
This is because to reproduce the current gas and total mass in the 
disk the NI models must have a lower SFR, leading to less D-consumption.  

Unfortunately, the NI model and those with Z$_{infall} \geq$ Z$_{\odot}$
do not satisfy several important observational constraints such as the
abundance gradients and the metallicity distribution of the G-dwarfs
(e.g. Tosi 1988b, MF).  Besides, it would be quite improbable that the
metallicity of the infalling material is higher than solar, since both 
the halo and the closest external galaxies (LMC and SMC) are metal poor 
compared to the Sun.  Thus, {\it standard} models with D-depleted 
infall fail to reconcile a high primordial abundance of deuterium 
with the observed solar system and ISM values.  

Next we study the effect of winds on the evolution of the Galaxy halo 
and disk using models based on the ``sequential" models described by MF 
and the ``two-infall" models of Chiappini, Matteucci \& Gratton (1997a, 
hereinafter CMG) which permit different choices of the star formation 
rate, of the IMF and of the relative importance of winds. 

\subsection{Models With Winds}

We have modified the ``sequential" models (MF) and the ``two-infall"
models (CMG) by allowing the Galaxy to lose mass through winds and we 
have explored the consequences for the destruction of deuterium, for 
the overall age-metallicity relation, for the variations of the ratios 
of abundances with the overall metallicity and for the metallicity
distribution in old stars.  Before presenting our results we briefly
describe the basic assumptions and parameters adopted for these models.

\subsubsection{Sequential Models}

In these models (see MF for details and specific parameter choices)
there is a continuity between the halo and disk evolution in the sense 
that the halo and the disk form as distinct phases of a unique process.  
The Galactic disk is approximated by several independent rings, each 2 
$kpc$ wide, without exchange of matter among them.  Continuous infall 
of gas ensures that the surface mass density in each ring increases 
with time.  The star formation rate depends on the total surface 
mass density and on the surface gas density.  The infall rate varies 
with galactocentric distance in such a way that the disk forms from 
the inside out.

\subsubsection{Two-Infall Models}

These models assume that the Galaxy formed as a result of two main 
infall episodes.  During the first one the halo and part of the thick 
disk formed (some thick disk stars seem to have been accreted, as shown 
by Gratton et al. 1997), whereas the second episode is responsible for 
the thin disk.  The timescale for the formation of the halo and thick 
disk is quite short ($\sim 0.5-1$ Gyr) whereas the timescale for the 
formation of the thin disk is quite long ($\sim 8$ Gyr) suggesting that 
the thin disk formed mainly out of extragalactic primordial gas.  This 
timescale ensures a very good fit to the new data on the G-dwarf 
metallicity distribution (Rocha-Pinto \& Maciel 1996, 1997).  The SFR 
is of the same form as in MF described above, except that a threshold 
in the surface gas density ($\sim 7 M_{\odot}pc^{-2}$) is assumed.  
When the gas density drops below this threshold, star formation stops.  
Such a threshold has been suggested by some star formation studies 
(Kennicutt 1989).  

Since the presence of a threshold leads naturally to a period between 
the first and the second infall episode where there is no active star 
formation, these models reproduce well the data in the most recent 
compilation by Gratton et al. (1997) which reveals the presence of a 
gap in star formation at the end of the thick-disk phase.  The main 
advantage of the two-infall models is the decoupling between the rate 
of gas loss from the halo and that of gas infalling onto the disk.  This 
allows a much longer timescale for disk evolution compared to that of 
the halo and thick disk phases.  An important aspect of these models 
is that the threshold adopted in the star formation process limits star 
formation in the halo phase, allowing for the formation of stars with 
low-metallicity, as required by the observed metallicity distribution 
of halo stars. 

\subsubsection{Ranges of Parameters}

For both the sequential models and the two-infall models there are 
many model parameters which may be varied.  In probing the influence 
of winds, we assume that the mass-loss is proportional to the SFR 
and we can choose to utilize such winds either in the disk, the halo, 
or both.  The relative importance of the winds is controlled by the 
ratio, w, of the mass-loss rate to the SFR (see Matteucci \& Tosi 1985) 
which we vary from 1.5 to 20.  Furthermore, the winds can expell all 
the elements or only those produced by massive stars.  Considering the 
production sites of the most important elements, preferential loss of 
gas ejected by massive stars implies removal of most of the new oxygen 
(which is only synthesized in massive stars), part of the carbon (which 
is produced in both high and intermediate mass stars), a small part of 
the iron (which is mostly produced by SNeI, with a small contribution 
from SNeII) and a negligible fraction of the nitrogen (which is produced 
almost exclusively by intermediate mass stars).  Since such a wind will 
modify the element ratios predicted for halo stars, it is crucial to 
compare the model predictions with the available observational data 
on the {\it relative} abundances of the elements.

For the timescale of the SFR in the halo we have considered models 
with 0.1 -- 0.2 Gyr, while for that in the disk we've allowed 1 -- 2 
Gyr.  For the formation timescale of the halo and for the disk we've 
explored 0.03 -- 8 Gyr.  We have allowed any infall to either be 
primordial ($\Z=0$) or enriched ($\Z=0.2\Z_{\odot}$).  Finally, 
we have experimented with several different IMFs (Scalo 1986; 
Scully et al. 1996, 1997), including  time-dependent cases 
(Padoan et al. 1997; Chiappini et al. 1997b; Scully et al. 1997).

We have run a very large number of models, both sequential and two-infall, 
varying all the above parameters in various combinations and we have 
identified the common characteristics as well as any significant 
differences.  Table 1 lists the adopted parameters for a significant
sub-sample of the examined cases.  In the next sections we choose three 
representative models to illustrate the behavior found for the sequential 
models and three for the two-infall models.

\section{Results}

Our results for D-destruction and chemical evolution are shown 
in Figures 3 -- 7.  The results for three representative sequential 
models (I, M, Q) are shown in Figures 3, 5 \& 7.  Model I uses the 
Padoan et al. (1997) IMF while model Q employs that of Scalo (1986).  
For model M, we adopt the metallicity-dependent IMF from Scully et al. 
(1996).  With these models it is easy to achieve significant, even 
excessive, D-destruction, as seen in Figure 3.  Indeed, many models 
of this type which we have investigated permit even more D-destruction 
than that shown for model I in Figure 3.  However, these models encounter 
severe problems which are evident in Figure 5 and in the lower panel 
of Figure 7.  Tuned to cycle gas through stars efficiently, so as to 
destroy primordial deuterium, these models predict excessive overabundances 
of oxygen and magnesium (relative to iron) at low metallicities.  The 
G-dwarf metallicity distributions predicted by these models are orthogonal 
to those observed: high at low metallicity where the data are low, low 
at high metallicity where the data are high.  

This inconsistency in the predicted G-dwarf distribution compared to that 
observed may be overcome if it is assumed, {\it ad hoc}, that the IMF during 
the earliest epochs (e.g., the first 1 Gyr, as in Scully et al. 1997) has 
no stars with mass below 2 M$_{\odot}$.  However, we have always found in 
such cases that if the SFR is consistent with the observational constraints, 
the predicted metallicity of the oldest stars is excessive (e.g., despite 
the ameliorating effect of the wind, oxygen reaches solar abundance in the 
first 1--2 Gyr).  If, instead, we constrain the metallicity to remain within 
the observational range, we are forced to adopt a very low SFR, at least a 
factor of ten lower than the minimum presently observed local rate.  

For the two-infall case we display the results of three representative 
models (A, C, F) in Figures 4, 6 \& 7.  These models all assume infall 
of unprocessed (primordial) material.  Models A \& C use the Scalo (1986) 
IMF while model F uses the Padoan et al. (1997) IMF.  For model C winds 
are very important; the ratio of the mass-loss rate to the SFR is w = 20.  
This is required in order to compensate for the overabundance of oxygen 
predicted by these models which all have a very high initial SFR.  As 
seen in Figure 4, in these models the deuterium is destroyed rapidly 
during the first phase but restored during the second infall episode.  
Thereafter, these two-infall models have only modest D-destruction.  
Starting with a high initial abundance of deuterium there is no way 
these models can account for the relatively low abundances observed 
in the solar system or the present local ISM.  In Figure 6 the predicted 
relative metallicities (O/Fe, Mg/Fe, C/Fe and N/Fe), as a function 
of [Fe/H], are compared with the data.  Despite the large spread in 
the data, models A, C, \& F provide a rather poor fit to the abundance 
ratios and are not in complete agreement with the observed G-dwarf 
metallicity distribution shown in the upper panel of Figure 7.  Although 
many two-infall models can be found which are consistent with the observed 
metallicities, all such models permit only modest D-destruction, by a 
factor which ranges from nearly unity (infall compensating for stellar 
destruction) up to $\sim 1.9$ at the time of the formation of the solar 
system and up to $\sim 2.2$ at the present epoch. 

The obstacle to obtaining large D-depletion is that it is generally 
necessary to assume much larger star formation efficiencies in the 
halo than in the disk, coupled with short timescales for the formation 
of both components.  It is apparent from the figures that only the 
sequential models, such as I, M and Q, are capable of sufficient (or 
even excessive) D-destruction.  This is because the second infall 
episode, which gives rise to the disk in the two-infall models, 
replenishes the ISM with large amounts of deuterium, thus compensating 
for most of its previous consumption.  The sequential models fail, 
however, in comparison with the other observational constraints.  In 
particular, their predicted G-dwarf metallicity distributions, shown 
in Figure~\ref{cf5}, are dramatically different from those observed.  
The reason is that in these models infall occurs only during halo and 
disk formation and the very short timescales for these phases permit only
very little gas accretion.  Consequently, the rapid consumption of the
available -- nearly pristine -- gas yields high numbers of metal-poor
stars.  Hence, these models lead to a predicted metallicity distribution
which is shifted towards low metallicities compared to what is observed.  

From our extensive study of all these models it is clear that those 
which destroy large amounts of deuterium do not fit the other constraints, 
whereas those which are in good agreement with the variety of observational 
data have much less deuterium consumption (depletion by factors smaller 
than 2--3).  Basically, the relatively high gas content and low metallicity 
of the local ISM is incompatible with efficient cycling of the gas through 
stars.  In contrast, Scully et al. (1997) claim their model II does fit the 
observational constraints while destroying deuterium by an order of magnitude.  
However, for the following reason their model II cannot be describing the 
chemical evolution of the solar vicinity of the Galaxy.  Scully et al. 
(1997) fix the overall normalization of their SFR by requiring that model II 
reproduce the observed present, local gas fraction.  They do not normalize
their SFR to the present, local SFR.  As a result, they failed to notice
that the predicted present SFR for model II is an order of magnitude 
smaller than that observed locally (Tinsley 1980, Timmes et al. 1995).  
If the model II SFR were renormalized to bring it into agreement with 
the observed local value, several observational obstacles would be 
encountered.  An increased SFR leads to more rapid production of the 
heavy elements, along with even more deuterium destruction.  To compensate 
for the former, perhaps the outflow rate could be increased as well.  
But this will result in a much smaller present gas fraction, no longer 
consistent with the data.  Furthermore, since the model II initial SFR 
is nearly 200 times higher than that at present, if the present SFR were 
normalized to that observed locally, the initial SFR would have been far 
too large to be consistent with the observed luminosity distribution of 
galaxies (Pozzetti, Bruzual \& Zamorani 1996; i.e., young galaxies would 
be too bright) and also inconsistent with the observed age-metallicity 
relation (Twarog 1980, Scalo 1986, Prantzos 1998).  To test these 
expectations we have constructed models R, S, T, and U (see Table 1) using 
the IMF of Scully et al. (1997) and employing a SFR and a variety of winds 
analogous to those in their model II.  In all of these cases we confirm 
that the metallicity is too high (reaching the solar value at $\sim 1-2$ 
Gyr) and the present gas fraction too small for consistency with the 
observational data.

\section{Discussion}

Observations in the local ISM and in the solar system provide valuable 
lower bounds to the primordial abundance of deuterium.  If these data 
are to be utilized to predict, or bound from above, the primordial 
abundance it is necessary to calculate, or bound, the destruction of 
deuterium in the Galaxy during its evolution.  Although most models of 
chemical evolution suggest only modest destruction (e.g., Audouze \& 
Tinsley 1974; Steigman \& Tosi 1992, 1995; Edmunds 1994; Galli et al. 
1995; Palla, Galli \& Silk 1995; Pilyugin \& Edmunds 1996; Prantzos 
1996), it is claimed that models may be designed to destroy a significant 
fraction of primordial deuterium (e.g., Vangioni-Flam \& Audouze 1988; 
Vangioni-Flam, Olive \& Prantzos 1994; Scully \& Olive 1995; Vangioni-Flam 
\& Cass\' e 1995; Scully et al. 1996, 1997).  To address this impasse we 
have first pursued an approach which is nearly independent of any chemical 
evolution models, relying only on the simple fact that to destroy 
deuterium gas must be cycled through stars and stars produce heavy 
elements.  We have found that destruction of D by more than a factor of 
five seems unlikely (see the upper panel of Figure 1).  However, we have 
noted that since elements such as oxygen are synthesized in massive stars 
while the bulk of the gas is cycled through lower mass stars, this 
constraint on D-destruction may be circumvented if the debris of the 
massive stars is removed from the disk of the Galaxy permanently (see the 
lower panel of Figure 1).  Alternatively, although a radically different 
early IMF might permit significant D-destruction while avoiding 
overproduction of the heavy elements, there is no evidence for any 
variations in the IMF (Wyse 1997).  Furthermore, such a radically 
different IMF would inevitably alter the abundance ratios of elements 
produced in different stellar sites (e.g., oxygen/iron, nitrogen/iron, 
etc.).  Recently Chiappini et al. (1997b) studied the effects of a variable 
IMF and concluded that although the IMF may be a function of time, such a 
time-variation must be small during the disk lifetime if the available 
observational constraints in the solar vicinity are to be satisfied.  
We have also considered the effect of infall of partially processed 
material in which some fraction of the primordial deuterium has been 
destroyed.  Even in these models D-destruction is limited to a factor 
of $\sim 1.4 - 2.0$ when the Sun formed and $\sim 1.6 - 2.3$ at present.

To explore the observational constraints on D-destruction further we 
have attempted to couple the evolution of the halo with that of the 
disk while allowing for high initial SFR and outflow (winds) along 
with infall.  Although the two-infall models appear consistent with 
the heavy element data, they permit only modest destruction: less 
than a factor of $\sim 1.9$ when the Sun formed and $\sim 2.2$ at 
present.  In contrast, the sequential halo/disk models allow almost 
any D-destruction factor, but none of the models we explored are 
consistent with the bulk of the observational data.  

Although there are no ``theorems" forbidding large D-destruction, 
it is quite difficult to break the connection between large gas 
fraction, low metallicity and modest D-destruction.  Star formation,
infall and outflow are linked inextricably in the ecology and
evolution of the Galaxy.  Rapid star formation is required if a large
fraction of the initial deuterium ($\gsim$ 90\%) is to be destroyed.  
But, rapid star formation results in rapid production of the heavy 
elements and in a rapidly decreasing gas fraction.  Infall of primordial 
or nearly unprocessed gas may ameliorate the effect of a high SFR by 
diluting the metallicity and replenishing the gas trapped in long-lived 
stars and stellar remnants.  However, the deuterium in the infalling 
gas drives the interstellar abundance back towards the higher primordial 
value resulting in less net D destruction.  Outflow of processed material 
may help keep the metallicity low, but when coupled with a large SFR, 
will result in a small gas fraction at present.  For these reasons it 
is difficult to design chemical evolution models which destroy deuterium 
efficiently while maintaining consistency with the other observational 
data.

In all the models that we and others have explored which are consistent
with the observational data, the D-destruction at present is limited 
to a factor of 3 or less.  If we adopt this as an upper bound, then 
we may bound from above the primordial abundance of deuterium: 
$X_{2\rm P} \leq 3X_{2\rm ISM}$.  Adopting the Linsky et al. (1993) 
value for (D/H)$_{\rm ISM} = 1.6 \pm 0.1 \times 10^{-5}$ and assuming 
for the hydrogen mass fraction at present $0.70 \pm 0.01$ and 
primordially $0.76 \pm 0.01$, we derive a $2\sigma$ upper bound of: 
(D/H)$_{\rm P} \leq 5.0 \times 10^{-5}$.  Clearly, this inferred 
upper bound to primordial deuterium is in conflict with the high D/H 
values claimed for some QSO absorbers (Carswell et al. 1994; Songaila 
et al. 1994; Rugers \& Hogan 1996) but is entirely consistent with the 
low values derived by Tytler et al. (1996) and by Burles \& Tytler 
(1996).  For standard BBN this upper bound to primordial D/H 
corresponds to a lower bound to the universal ratio of nucleons-to-photons 
$\eta_{10} \geq 4.0$ or, in terms of the baryon density parameter 
($\Omega_{\rm B}$) and the Hubble parameter (H$_{0} = 100h~km/s/Mpc$): 
$\Omega_{\rm B}h^{2} \geq 0.015$.  Such a lower bound to $\eta_{10}$, 
in the context of standard BBN, leads to a predicted lower bound on 
the primordial mass fraction of helium-4 (Y$_{\rm P} \geq 0.244$) 
which is in modest disagreement with that inferred from observations 
of low-metallicity extragalactic \hii regions (Olive \& Steigman 
1995; Olive, Skillman \& Steigman 1997) unless systematic errors 
are responsible for the low value of Y$_{\rm P}$ derived from the 
data.  This latter possibility receives support from the new data 
of Thuan \& Izotov (1998) who derive Y$_{\rm P} = 0.244 \pm 0.002$.  
Primordial deuterium is an invaluable baryometer.  Until the 
observational situation of nearly primordial gas in QSO absorbers is 
resolved by more data, local ISM and/or solar system data in concert 
with estimates or bounds to Galactic D-destruction will continue to 
provide useful cosmological constraints.

\begin{acknowledgments}
We are pleased to thank our referee, Bernard Pagel, for important
suggestions which have led to improvements in our discussion.  G. S. 
thanks Keith Olive and Sean Scully for revealing discussions of their 
chemical evolution models.  The work of G. Steigman at OSU is supported 
by a research grant from the DOE.  The work of C. Chiappini is supported 
by FAPESP in Brasil and by ICTP in Italy.  Funds from the Italian CNR 
through a GNA contract are also acknowledged.  C. C. and F. M. thank 
SISSA for hospitality. 
\end{acknowledgments}

\clearpage

\clearpage

\begin{deluxetable}{cccccccccc}
\footnotesize
\tablecaption{Model Parameters}
\tablewidth{0pt}
\tablehead{
\colhead{Model}  &
\colhead{type} &
\colhead{wind$^{(a)}$} &
\colhead{w} &
\colhead{$ \stackrel{\textstyle \nu_h }{ ({\rm Gyr}^{-1}) } $} & 
\colhead{$ \stackrel{\textstyle \nu_d }{ ({\rm Gyr}^{-1}) } $} & 
\colhead{$ \stackrel{\textstyle \tau_h }{ ({\rm Gyr}) } $} & 
\colhead{$ \stackrel{\textstyle \tau_d }{ ({\rm Gyr}) } $} & 
\colhead{infall$^{(b)}$} & 
\colhead{IMF$^{(c)}$} 
}
\tablenotetext{(a)}{D = Differential wind (removes SN ejecta); N = Normal 
wind (removes SN ejecta and local gas)}
\tablenotetext{(b)}{P = Primordial; E = Enriched with $Z_{inf}=0.2Z_{\odot}$}
\tablenotetext{(c)}{S86 = Scalo 1986; P97 = Padoan et al. 1997; S96 = Scully 
 et al. 1996; Scully et al. 1997}

\startdata

A &2-infall & D; halo &1.5 & 10 & 1.0 & 0.03 & 8.0 & P & S86 \nl
B &2-infall & D; halo &1.5 & 10 & 1.0 & 0.03 & 8.0 & E & S86 \nl
C &2-infall & N; halo & 20. & 10 & 1.0 & 0.03 & 8.0 & P & S86 \nl
D &2-infall & D; halo + disk & 1.5 & 10 & 1.0 & 0.03 & 8.0 & P & S86 \nl
E &2-infall & N; halo + disk & 20. & 10 & 10 & 0.03 & 8.0 & P & S86 \nl
F &2-infall & D; halo & 1.5 & 10 & 1.0 & 0.03 & 8.0 & P & P97 \nl
G &sequential & D; halo & 1.5 & 10 & 10 & 0.03 & 0.03 & P & S86 \nl
H &sequential & D; halo & 1.5 & 10 & 0.5 & 0.03 & 0.03 & P & S86 \nl
I &sequential & D; halo & 1.5 & 10 & 0.5 & 0.03 & 0.03 & P & P97 \nl
M &sequential & D; halo & 1.5 & 10 & 0.5 & 0.03 & 0.03 & P & S96 \nl
N &sequential & D; halo & 1.5 & 10 & 0.5 & 8.0 & 8.0 & P & S86 \nl
O &sequential & D; halo & 1.5 & 10 & 0.5 & 3.0 & 3.0 & P & S86 \nl
P &sequential & D; halo & 1.5 & 10 & 0.5 & 0.5 & 0.5 & P & S86 \nl
Q &sequential & D; halo & 1.5 & 5 & 0.5 & 0.03 & 0.03 & P & S86 \nl
R &sequential & D; halo & 1.5 & 0.5 & 0.5 & 0.03 & 0.03 & P & S97 \nl
S &sequential & N; halo + disk & 1.5 & 0.5 & 0.5 & 0.03 & 0.03 & P & S97 \nl
T &sequential & N; halo + disk & 10 & 0.5 & 0.5 & 0.03 & 0.03 & P & S97 \nl
U &sequential & D; halo + disk & 1.5 & 0.5 & 0.5 & 0.03 & 0.03 & P & S97 \nl

\enddata
\end{deluxetable}

%
%

\clearpage

\figcaption{Upper limits to the virgin fraction of the ISM, see eq.(5), 
resulting from the oxygen yields weighted over Tinsley's (1980) IMF. 
Solid lines correspond to CC79 yields and dotted lines to WW95 yields. 
The top panel shows the dependence on the IMF lower mass limit (with 
$m_u$=100\msun), the bottom panel the dependence on the upper mass limit 
(with $m_l$=0.8 or 5 \msun).
\label{imf}}

\figcaption{Evolution in the solar ring of the D abundance by mass.
 Vertical bars give the 2$\sigma$ ranges for the abundances derived from
solar system and local ISM observations. The dotted curve and the solid
one correspond to DST's models NI-Z-III and 1-T-Vb, all the others to 
models assuming depleted D in the infalling gas (X$_{2,infall}=0$ for the
short-dashed line, X$_{2,infall}=0.1X_{2P}$ for the long-dashed line,
and X$_{2,infall}=0.2X_{2P}$ for the dash-dotted line).
\label{dinf}}

\figcaption{Same as Figure~2 for the sequential models I (dot-dashed line),
M (solid line) and Q (dashed line).
\label{cf1}}

\figcaption{Same as Figure~3 for the two-infall models A (solid line),
C (dashed line) and F (dot-dashed line).
\label{cf2}}

\figcaption{The four panels show the relative abundances [O/Fe], [Mg/Fe], 
[C/Fe] and [N/Fe] versus [Fe/H] predicted for the sequential models.  
The models are labeled as in Figure~\ref{cf1}. The data were taken from 
Laird (1985), Gratton and Ortolani (1986), Tomkin et al. (1986), Carbon 
et al. (1987), Gratton and Sneden (1987), Magain (1987, 1989), Edvardsson 
et al. (1993) and Gratton et al. (1997).
\label{cf3}}
 
\figcaption{Same as Figure 5 for the two-infall models.  The models are 
 labeled as in Figure~\ref{cf2}.
\label{cf4}}

\newpage

\figcaption{G-dwarf metallicity distribution in the solar vicinity predicted
 by the different models. The upper panel is for the two-infall models and 
 the lower panel is for the sequential models.  The data are from Rocha-Pinto 
 and Maciel (1996).  The models are labeled as in Figures \ref{cf1} \&  
 \ref{cf2}.
\label{cf5}}


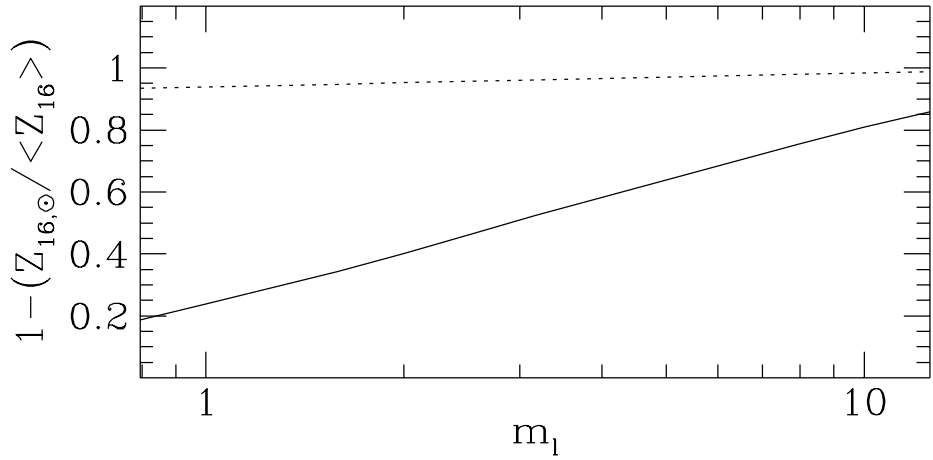
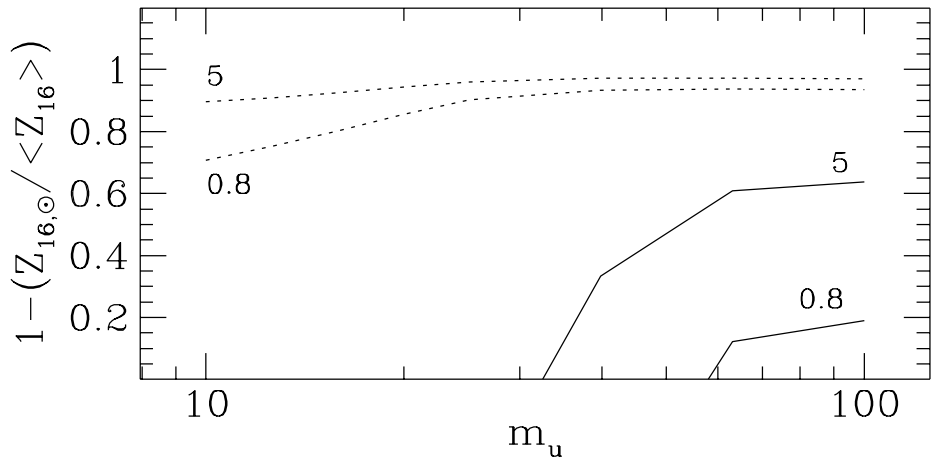
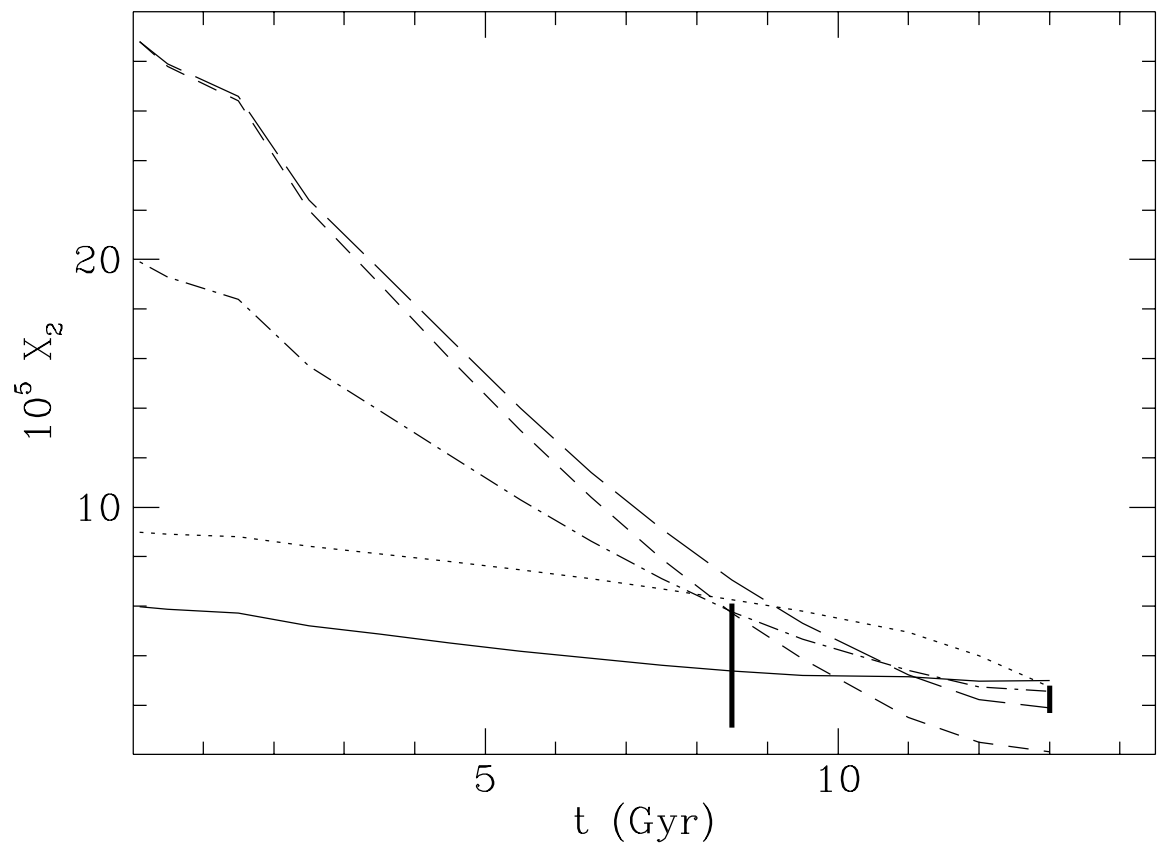
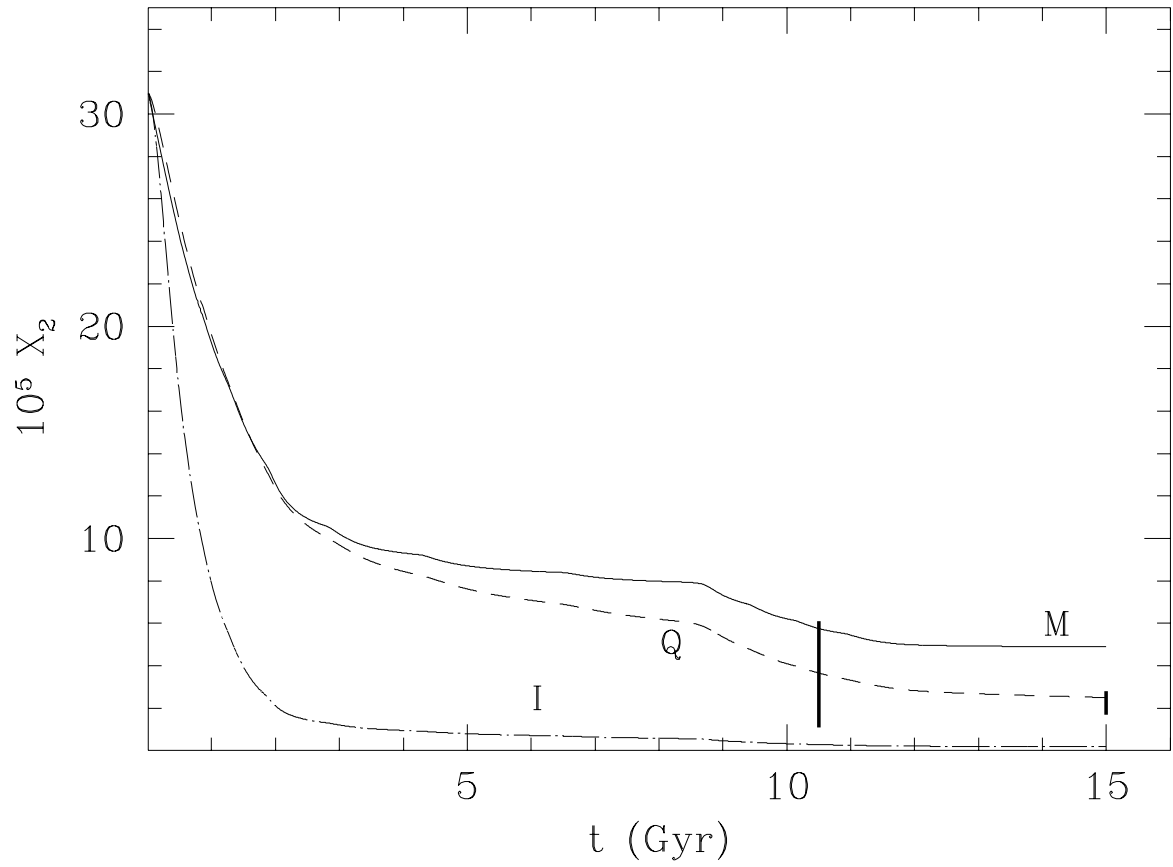
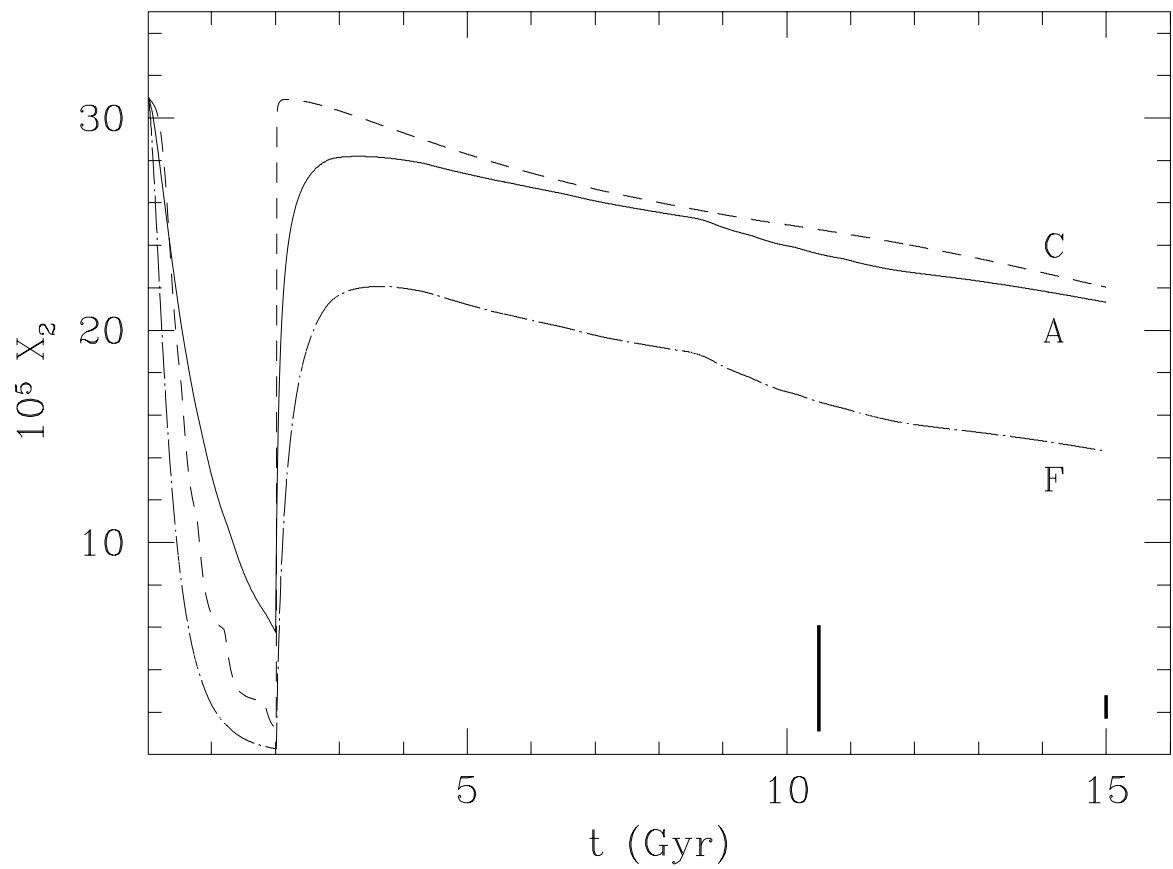
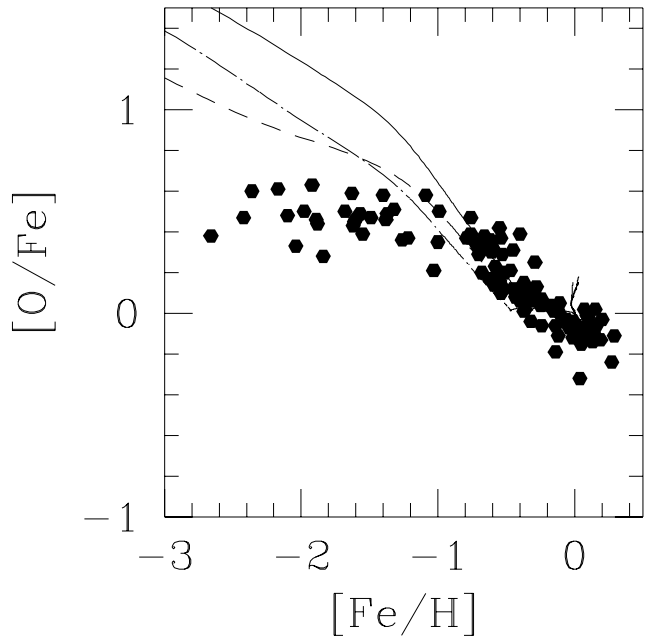
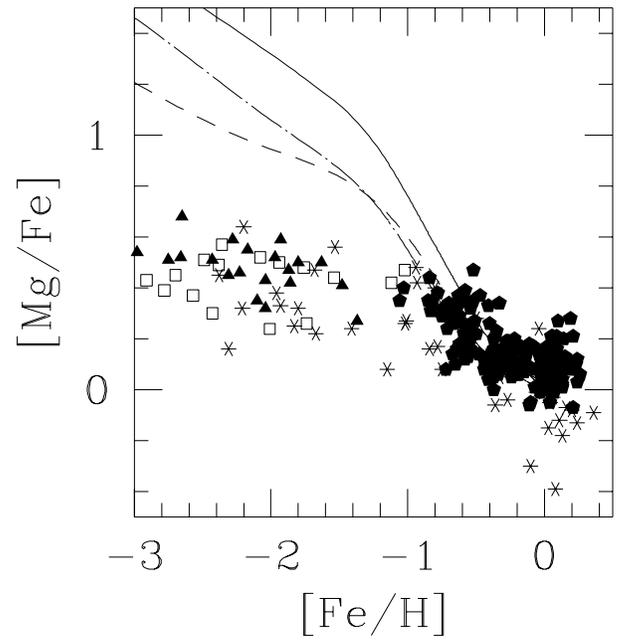
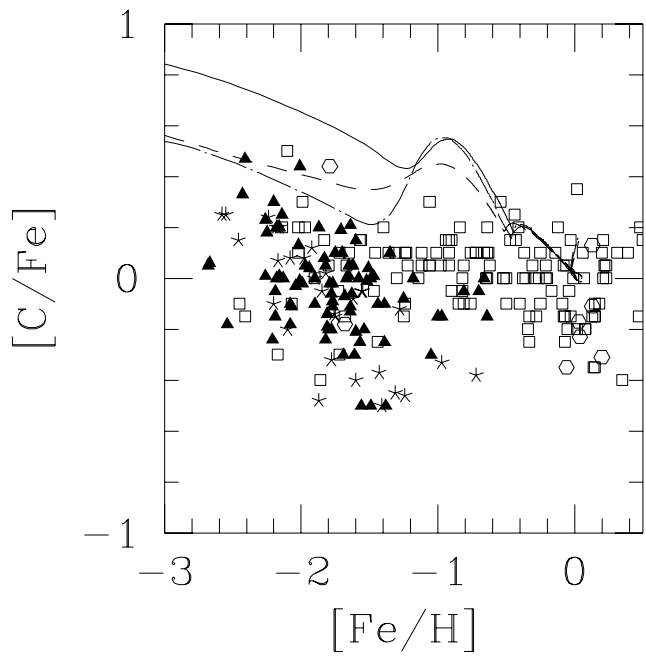
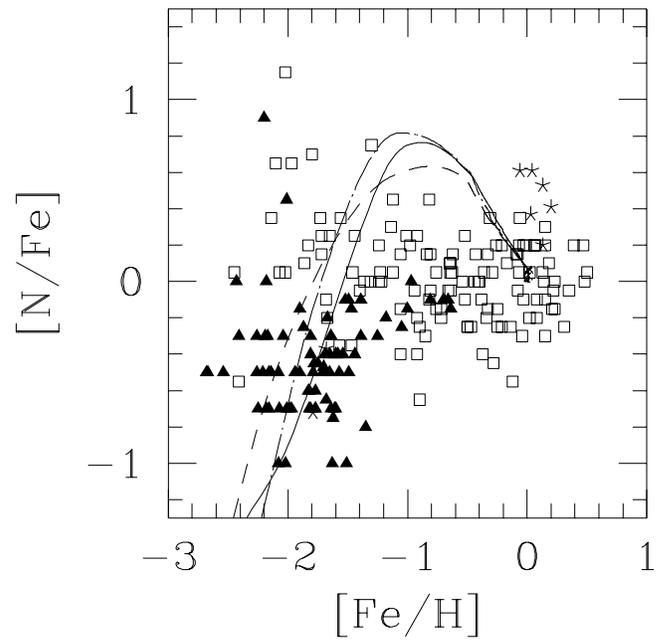
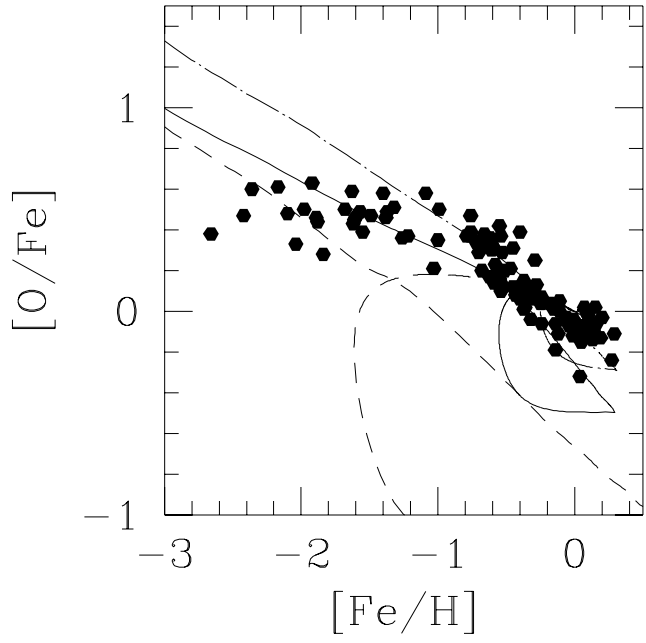
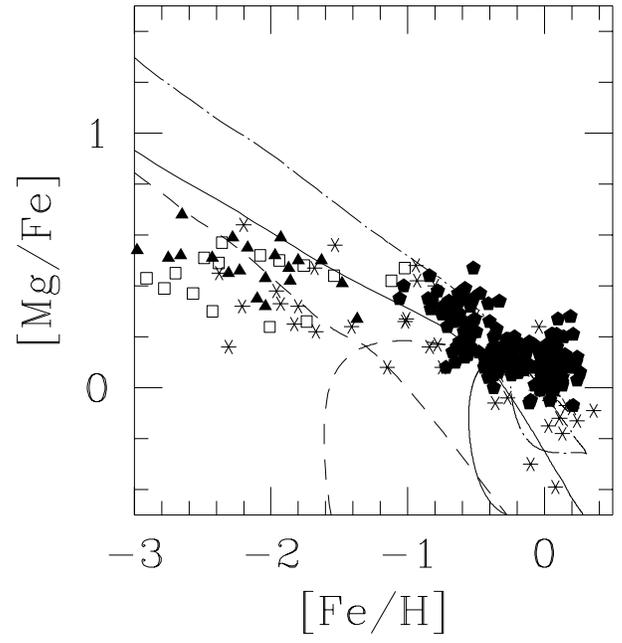
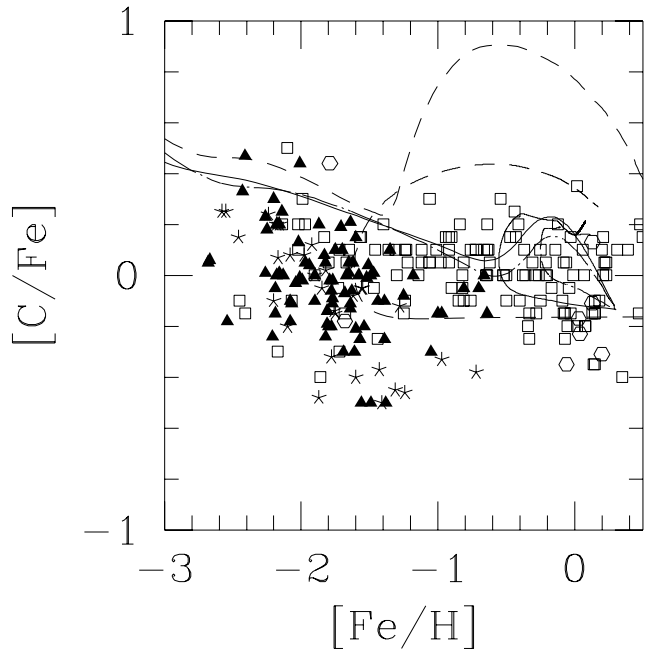
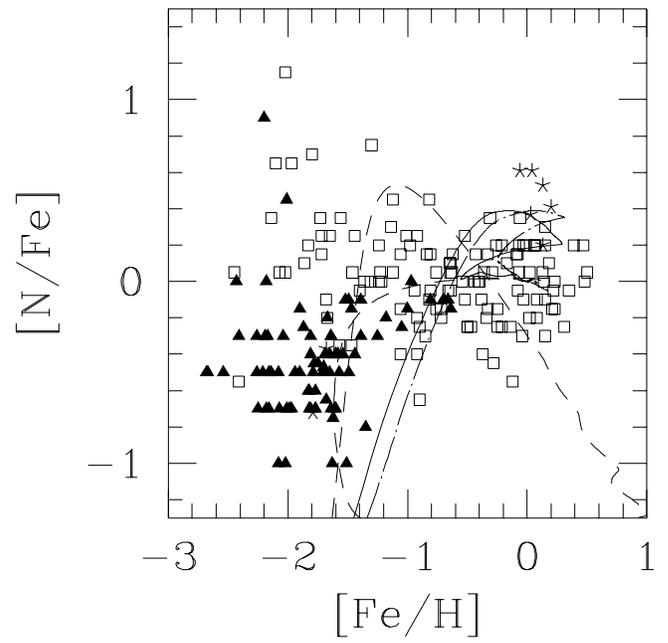
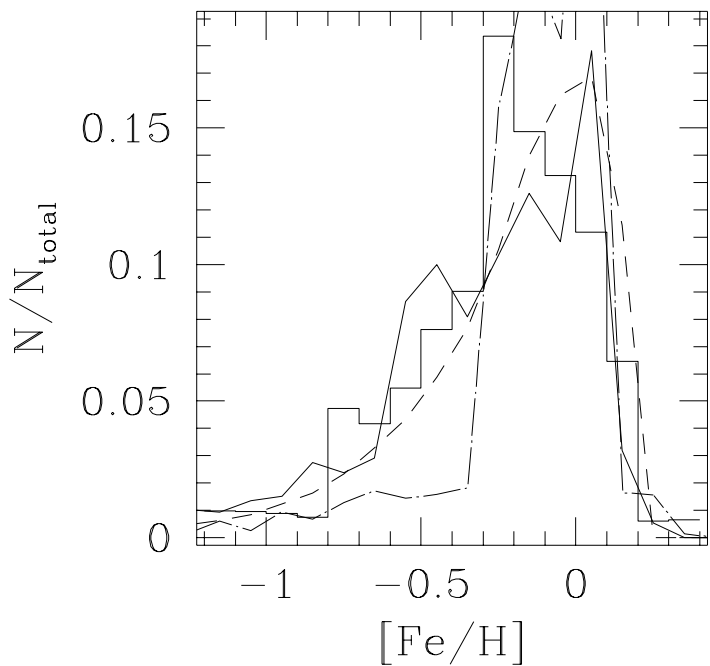
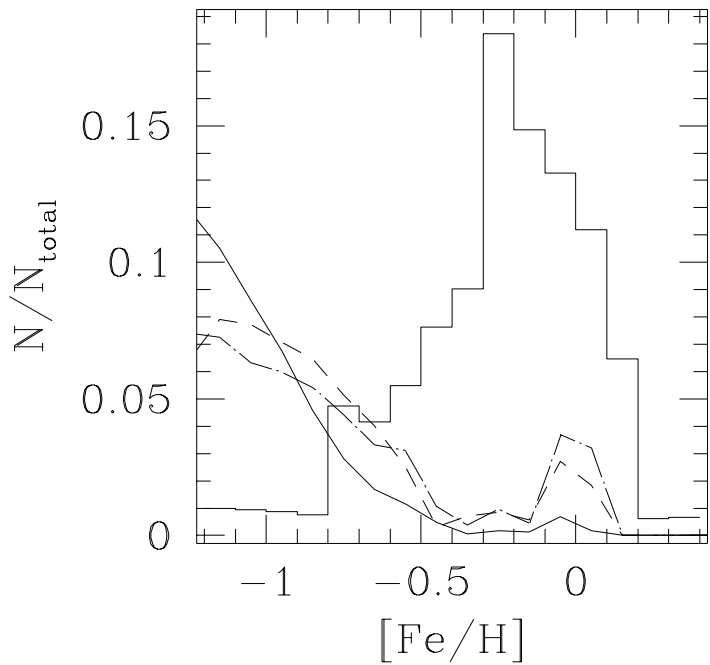
\end{document}